\documentclass[twocolumn]{aastex63}
\usepackage{verbatim}
\usepackage{comment}

\makeatletter
\def\@hex@@Hex#1%
 {\if a#1A\else \if b#1B\else \if c#1C\else \if d#1D\else
  \if e#1E\else \if f#1F\else #1\fi\fi\fi\fi\fi\fi \@hex@Hex}
\makeatother
\definecolor{afcolor}{HTML}{b3443c}

\definecolor{apcolor}{HTML}{b3003b}

\shorttitle{Super-early luminous galaxies}
\shortauthors{Ferrara et al.}
\graphicspath{{./}{figures/}}

\begin{document}



\def\be{\begin{equation}}
\def\ee{\end{equation}}
\newcommand\code[1]{\textsc{\MakeLowercase{#1}}}
\newcommand\quotesingle[1]{`{#1}'}
\newcommand\quotes[1]{``{#1}"}
\def\gsim{\lower.5ex\hbox{\gtsima}} 
\def\lsim{\lower.5ex\hbox{\ltsima}} 
\def\gtsima{$\; \buildrel > \over \sim \;$} 
\def\ltsima{$\; \buildrel < \over \sim \;$} \def\gsim{\lower.5ex\hbox{\gtsima}} 
\def\lsim{\lower.5ex\hbox{\ltsima}} 
\def\simgt{\lower.5ex\hbox{\gtsima}} 
\def\simlt{\lower.5ex\hbox{\ltsima}}

\def\msun{{\rm M}_{\odot}}
\def\lsun{{\rm L}_{\odot}}
\def\dsun{{\cal D}_{\odot}}
\def\fsun{\xi_{\odot}}
\def\zsun{{\rm Z}_{\odot}}
\def\msunyr{\msun {\rm yr}^{-1}}
\def\gdens{\msun\,{\rm kpc}^{-2}}
\def\sfrdens{\msun\,{\rm yr}^{-1}\,{\rm kpc}^{-2}}

\def\mum{\mu {\rm m}}
\newcommand{\angstrom}{\mbox{\normalfont\AA}}
\def\cc{\rm cm^{-3}}
\def\uflux{{\rm erg}\,{\rm s}^{-1} {\rm cm}^{-2} }

\def\fdust{\xi_{d}}
\def\fesc{f_{\rm esc}\,}
\def\td{\tau_{sd}}
\def\Sg{$\Sigma_{g}$}
\def\S*{$\Sigma_{\rm SFR}$}
\def\Ssfr{\Sigma_{\rm SFR}}
\def\Sgas{\Sigma_{\rm g}}
\def\Sstar{\Sigma_{\rm *}}
\def\Sesc{\Sigma_{\rm esc}}
\def\Srad{\Sigma_{\rm rad}}

\def\Dsolar{${\cal D}/\dsun$}
\def\Zsolar{$Z/\zsun$}
\def\DDsolar{\left( {{\cal D}\over \dsun} \right)}
\def\ZZsolar{\left( {Z \over \zsun} \right)}
\def\kms{{\rm km\,s}^{-1}}
\def\skms{$\sigma_{\rm kms}\,$}

\def\Scii{$\Sigma_{\rm [CII]}$}
\def\Sciimax{$\Sigma_{\rm [CII]}^{\rm max}$}
\def\CII{\hbox{[C~$\scriptstyle\rm II $]~}}
\def\CIII{\hbox{C~$\scriptstyle\rm III $]~}}
\def\OII{\hbox{[O~$\scriptstyle\rm II $]~}}
\def\OIII{\hbox{[O~$\scriptstyle\rm III $]~}}
\def\HH{\hbox{H$_2$}~} 
\def\HI{\hbox{H~$\scriptstyle\rm I\ $}} 
\def\HII{\hbox{H~$\scriptstyle\rm II\ $}} 
\def\CIion{\hbox{C~$\scriptstyle\rm I $~}}
\def\CIIion{\hbox{C~$\scriptstyle\rm II $~}}
\def\CIIIion{\hbox{C~$\scriptstyle\rm III $~}}
\def\CIVion{\hbox{C~$\scriptstyle\rm IV $~}}
\def\nhh{n_{\rm H2}}
\def\nhi{n_{\rm HI}}
\def\nhii{n_{\rm HII}}
\def\fhh{x_{\rm H2}}
\def\fhi{x_{\rm HI}}
\def\fhii{x_{\rm HII}}
\def\fd{f^*_{\rm diss}} 
\def\ks{\kappa_{\rm s}}

\def\cyan{\color{cyan}}
\definecolor{apcolor}{HTML}{b3003b}
\definecolor{afcolor}{HTML}{800080}
\definecolor{lvcolor}{HTML}{DF7401}
\definecolor{mdcolor}{HTML}{01abdf} 
\definecolor{cbcolor}{HTML}{ff0000}
\definecolor{sccolor}{HTML}{cc5500} 
\definecolor{sgcolor}{HTML}{00cc7a}

\title{Super-early JWST galaxies, outflows and Ly$\alpha$ visibility in the EoR}

\correspondingauthor{Andrea Ferrara}
\email{andrea.ferrara@sns.it}

\author[0000-0002-9400-7312]{Andrea Ferrara}
\affil{Scuola Normale Superiore,  Piazza dei Cavalieri 7, 50126 Pisa, Italy}



\begin{abstract}
The overabundance of super-early (redshift $z>10$), luminous ($M_{\rm UV} < -20$), and blue galaxies detected by {\it JWST} has been explained \citep{Ferrara23} as due to negligible dust attenuation in these systems.  We show that such model correctly reproduces the UV luminosity function at $z>10$, and the star formation rate (SFR) density evolution. The model also predicts, in agreement with data, that the cosmic specific SFR grows as ${\rm sSFR} \propto (1+z)^{3/2}$. At $z \simeq 10$ the cosmic sSFR crosses the  critical value $\rm sSFR^\star = 25\, \rm Gyr^{-1}$ and  $\approx 45$\% of the galaxies become super-Eddington driving outflows reaching velocities of  $\approx 830 (\epsilon_\star/f_M)^{1/2}\ \kms$, where $\epsilon_\star$ and $f_M$ are the SF efficiency and fraction of the halo gas expelled in the outflow, respectively. This prediction is consistent with the outflow velocities measured in 12 super-Eddington galaxies of the {\it JWST}/JADES sample.  Such outflows clear the dust, thus boosting the galaxy luminosity. They also dramatically enhance the visibility of the Ly$\alpha$ line from $z>10$ galaxies, by introducing a velocity offset.  The observed Ly$\alpha$ properties in GN-z11 ($z=10.6$) are simultaneously recovered by the outflow model if $\log N_{\rm HI} \simeq 20.1$, implying that the outflow is largely ionized. We make analogous predictions for the Ly$\alpha$ visibility of other super-early galaxies, and compare the model with Ly$\alpha$ surveys at $z>7$, finding that essentially all super-Eddington (sub-Eddington) galaxies are (not) detected in Ly$\alpha$. Finally, the sSFR positively correlates with the LyC escape fraction as outflows carve ionized, transparent channels through which LyC photons leak. 
\end{abstract}


\section{Introduction} \label{sec:intro}
Among the many tantalizing mysteries the \textit{James Webb Space Telescope} ({\it JWST}) promises to unravel, one of the most captivating is the discovery of {super-early} (redshift $z \simgt 10$)  galaxies. These galaxies, often endearingly referred to as \quotes{Blue Monsters} \citep{Ferrara23, Ziparo23, Fiore23}, show a number of peculiar properties. They are (a) very bright ($M_{\rm UV} < -20$), (b) more numerous than expected (number density $\approx 10^{-5}\ \rm Mpc^{-3}$), (c) very blue (UV spectral slopes $\beta \simlt -2.0$), and (d) very compact (size $\approx 100$ pc).  

These enigmatic galactic giants represent a critical juncture in the cosmic timeline, bridging the gap between the Big Bang and the formation of the galaxies we see today. The study of super-early galaxies holds the potential to reshape our understanding of the universe's infancy, shedding light on the processes that catalyzed the birth of first stars and black holes, and the emergence of the structures that underpin cosmic tapestry.

After the powerful initial detection boost provided by the public {\it JWST} Early Release Science programs \citep{Naidu22, Donnan22, Finkelstein22, Atek22, Borsani22, Castellano22, Santini22, Adams22, Harikane2022, topping2022, rodighiero2022, bradley2022, barrufet2022,  trussler2022, leethochawalit2022, whitler2022, Austin23, McLeod23, Harikane23, Robertson23, Fujimoto23, Castellano23} efforts have mostly concentrated on obtaining spectroscopic confirmations of the photometric candidates. These observationally intensive studies \citep{Curtis23, Bunker23, Hsiao23, Haro23b, Bradac23, Wang23, Stiavelli23} have not only confirmed the photometric redshifts for the majority of the targets (although with noticeable exceptions, see \citealt{Haro23}); they have also provided key physical insights on the nature of these systems. 


The observed large number density of super-early galaxies poses a serious challenge to essentially all galaxy evolution models \citep{Mason22, Mirocha23, DiCesare23, Gong23, Furlanetto23, Kannan22, Yajima22, Keller22, McCaffrey23, Mauherhofer23, Pad23, Munoz23}, and perhaps to the $\Lambda$CDM one itself \citep{Boylan23, Lovell23, Haslbauer22, Steinhardt23}. Barring modified cosmological scenarios, in order to increase the UV luminosity of a galaxy population at fixed cosmic number density, there are essentially three possibilities. 

Recalling that the UV luminosity of a galaxy is $L_{\rm UV} \propto \rm SFR\, \kappa_{1500}\, e^{-\tau}$, where $\kappa_{1500}$ is the conversion factor from star formation rate (SFR) to luminosity, and $\tau$ is the UV dust optical depth, we can (a) increase the SFR,  (b) increase $\kappa_{1500}$, or (c) decrease $\tau$. The SFR can be increased if a larger efficiency of gas conversion into stars is assumed \citep{Dekel23, Renzini23, Sipple23}, or if SFR is stochastic \citep{Mirocha23, Shen23, Pallottini23}. Both solutions are unpalatable as sometimes efficiencies $>100$\% would be required, and because the r.m.s. of stochastic SFR variations is too small to explain the data \citep{Pallottini23, Ciesla23}. Alternatively, one could postulate the presence of a metal-free (PopIII) stellar population with a larger $\kappa_{1500}$ value \citep{Riaz22, Wang22, Maiolino23}. Although this cannot be excluded, the absence of prominent HeII lines in spectroscopically confirmed $z>10$ galaxies, together with their moderately high metallicities ($Z \simgt 0.1 \zsun$) make this hypothesis unlikely. Similarly, a significantly top-heavy IMF may partially, but not completely, alleviate the problem \citep{Trinca23}. 

Option (c) can be achieved in two ways. The first is to invoke a spatial segregation of stellar (optical-to-UV) and dust continuum (infrared) emitting regions \citep{Behrens18,Sommovigo21, Sommovigo22,Ferrara22a,Dayal22}. In this scenario, the UV radiation mostly comes from the transparent diffuse interstellar medium (ISM), hosting either little or cold dust. The dust-obscured SFR is instead located in giant molecular clouds, strongly emitting at IR wavelengths. This hypothesis can be excluded on the basis of the ALMA \citep{Bakx22, Popping22, Kaasinen22, Yoon22, Fujimoto22} and NOEMA \citep{Fudamoto23} far-infrared dust continuum non-detections \citep{Ziparo23}. 

Alternatively, given their typical stellar mass ($M_\star = 10^9\,\msun$) and dust-to stellar ratio ($\approx 1/1000$, \citealt[][]{Dayal22}), one would expect $10^6\ \msun$ of dust to be present in these systems. As we show below (Sec. \ref{Sec:outflows}) this dust amount would totally obscure the UV light emitted by the galaxy. Thus, some mechanism must expel or at least lift the dust off the galaxy main body. We suggest that radiation-driven outflows, developing once the galaxy luminosity becomes super-Eddington, might do the job \citep[][see also \citealt{Tsuna23}]{Ferrara23}. 

In this paper, following-up previous work in \citet{Ferrara23}, we study the conditions for early galaxies to develop radiation-driven outflows which increase the galaxy luminosity and produce their steep UV spectra. 

Thanks to the growing body of {\it JWST} observations, it is already possible to test this scenario up to the highest redshifts. The tests include, among others, the comparison with UV luminosity functions, individual and cosmic star formation rates, stellar ages and outflow properties deduced from spectral line information. Once validated, the model can be used to make novel predictions related to the visibility and spectral properties of the Ly$\alpha$ emission from galaxies in the Epoch of Reionization \citep[e.g.][]{Tang23, Endsley23, Stark17}. It might also provide information on the escape fraction of LyC photons, a crucial parameter to model the reionization of the intergalactic medium.

The paper is organized as follows\footnote{Throughout the paper, we assume a flat Universe with the following cosmological parameters: $\Omega_{\rm M} = 0.3075$, $\Omega_{\Lambda} = 1- \Omega_{\rm M}$, and $\Omega_{\rm b} = 0.0486$,  $h=0.6774$, $\sigma_8=0.826$, where $\Omega_{M}$, $\Omega_{\Lambda}$, and $\Omega_{b}$ are the total matter, vacuum, and baryon densities, in units of the critical density; $h$ is the Hubble constant in units of $100\,\kms$, and $\sigma_8$ is the late-time fluctuation amplitude parameter \citep{planck:2015}.}. In Sec. \ref{sec:super-Edd_gals} we derive a specific SFR-based  condition for a galaxy to develop a radiation-driven outflow; Sec. \ref{sec:sSFR} presents the model predictions for the sSFR evolution and various tests of the model using available {\it JWST} data. Sec. \ref{Sec:outflows} discusses the properties of the outflows and compares them with data from the {\it JWST}/JADES survey. Predictions for the Ly$\alpha$ visibility in the EoR  are given in Sec.  \ref{sec:Lya}, and are followed by a general discussion in \ref{sec:discussion}. A brief summary \ref{sec:summary} concludes the paper.   
%
%
\begin{figure*}[t]
\begin{tabular}{lr}
\includegraphics[width = 0.48 \textwidth]{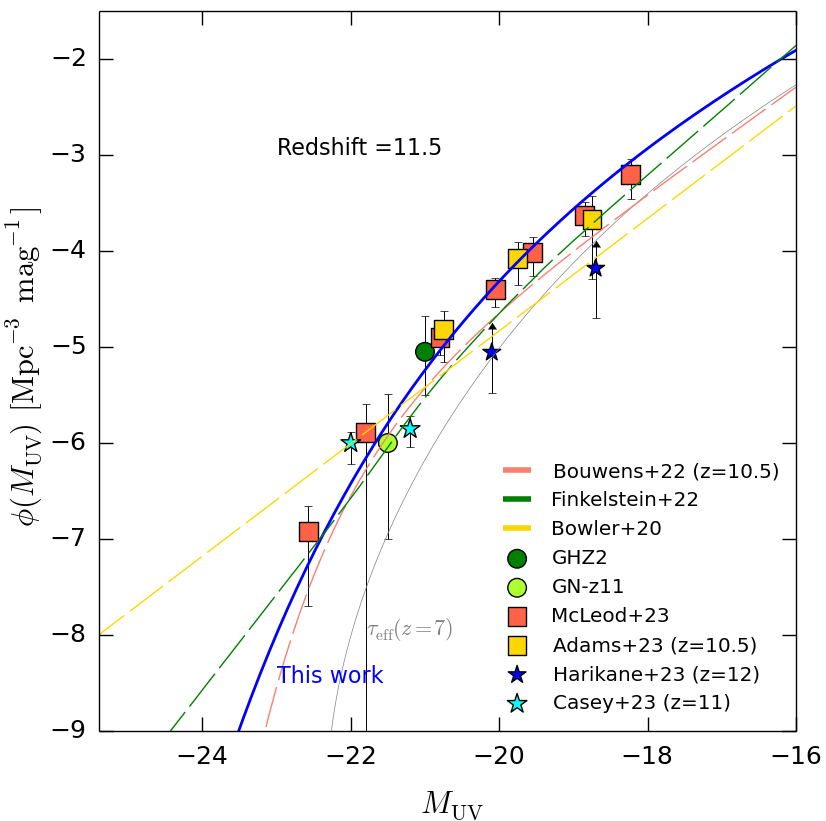}
\includegraphics[width = 0.475 \textwidth]{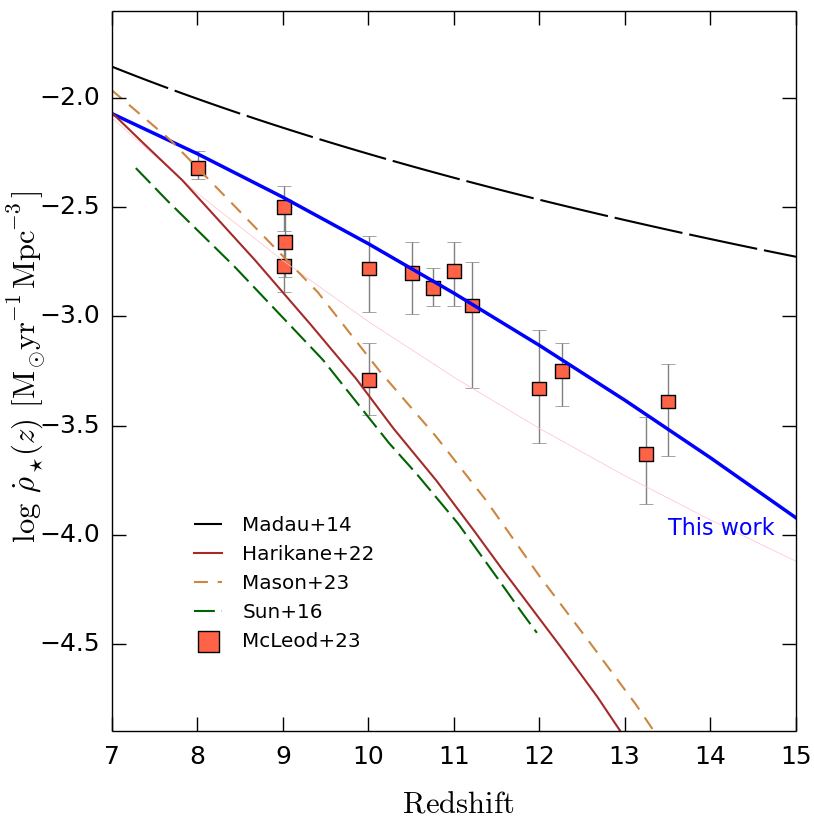}
\end{tabular}
\caption{Results of the \citet{Ferrara23} model used here. \textit{Left panel}: Predicted UV luminosity function (blue curve) compared with available data at $z \simeq 11.5$ (points). The blue curve assumes no dust-attenuation, while the grey curve marked as $\tau_{\rm eff}(z=7)$ assumes that the dust attenuation at $z=11.5$ is the one calibrated at $z=7$ using the ALMA REBELS data \citep{Bouwens22a, Ferrara22a}.   The data are from \citet{Castellano22} (GHZ2), \citet{Bunker23} (GN-z11), \citet{McLeod23}, \citet{Adams2023}, and \citet[][lower limits]{Harikane23}. Also shown are empirical fits provided by \citet[][red dashed]{Bouwens22b}, \citet[][yellow dashed]{Bowler20}, and \citet[][green dashed]{Finkelstein22}. \textit{Right:} Predicted evolution of the cosmic star formation rate density (blue curve), integrated down to a limiting AB mag $M_{1500} = -17$, compared with the data compilation of early {\it JWST} measurements in \citet{McLeod23}, also including data from \citet{Harikane23c, Oesch18, Donnan23a, Donnan23b, McLeod16, Perez23}.}
\label{Fig:model_test}
\end{figure*}
\section{Super-Eddington galaxies}\label{sec:super-Edd_gals}
We start by analyzing the conditions for which high-redshift galaxies might develop a radiatively-driven dusty outflow. To this aim we make two basic assumptions: (a) the stellar component is sufficiently compact that it can be treated as a point source; (b) gravity in the central regions is dominated by stars, so that the contribution of gas (and dark matter) can be neglected. The first hypothesis is justified by the very small sizes deduced for $z>8$ sources \citep{Tacchella23, Robertson23}.  The second is supported by the high typical stellar densities of these systems $\approx 1000\, \msun \rm pc^{-3}$ \citep{Charbonnel23} which might indicate that most of the gas in these regions has been converted into stars \citep{Dekel23}. 

We also assume that radiation pressure from the stars acts on dust \citep{fabian2006} which is tightly coupled by viscous and Coulomb drag forces to the gas, to which it transfers the radiation momentum. This implies that the classical Eddington luminosity $L_{E}=4\pi G m_p c M_*/\sigma_T = 1.26\times 10^{38}(M_*/M_\odot)\, \rm erg\, s^{-1}$ is reduced by a boost factor $A = \sigma_d/\sigma_T$, the ratio of the dust to Thomson cross-section.
Depending on the stellar age, metallicity, dust-to-gas ratio, and radiative transfer effects the boost factor can be computed from the \code{CLOUDY} photoionization code. \citet{Fiore23} found it to be in the range $A=100-600$. Such range brackets uncertainties in the above effects (see Fig. 2 of \citealt[][]{Fiore23}). Here, we conservatively use $A=100$. 

The super-Eddington condition
\begin{equation}
L_{\rm bol} > L_E^{\rm eff} = A^{-1} L_E, 
\label{eq:superEdd}
\end{equation}
where $L_{\rm bol}$ is the source bolometric luminosity, can be equivalently written as $\lambda_E^{\rm eff} > 1$, where the (effective) Eddington parameter $\lambda_E^{\rm eff}= L_{\rm bol}/L_E^{\rm eff}$ has been introduced. If the galaxy becomes super-Eddington, a radiatively-driven outflow will develop, ejecting both dust and gas from the system. 

We determine the galaxy (unattenuated) UV luminosity at 1500\AA, $L_{1500}$, from the star formation rate (SFR) via a conversion factor, ${\cal K}_{1500}$ $[{L_\odot}/M_\odot {\rm yr}^{-1}]$, whose value has been chosen so to match the one used by the ALMA REBELS survey \citep{Bouwens22a}: ${\cal K}_{1500} \equiv {L_{1500}}/{\rm SFR} = 0.587 \times 10^{10}$. We then convert the UV luminosity into $L_{\rm bol}$ using a bolometric correction, $f_{\rm bol} = L_{\rm bol}/L_{1500} = 2$, in agreement with the galaxy templates used to evaluate the boost factor \citep{Fiore23}.

The super-Eddington condition, $\lambda_E^{\rm eff} > 1$, with the above definitions, translates into one on the specific star formation rate, ${\rm sSFR=SFR}/M_\star$: 
\begin{equation}
{\rm sSFR} > {\rm sSFR}^\star \simeq 25 \left(\frac{100}{A}\right) \left(\frac{2}{f_{\rm bol}}\right) {\rm Gyr^{-1}}.
\label{eq:ssfr_thresh}
\end{equation}
Eq. \ref{eq:ssfr_thresh} represents the necessary condition for a galaxy to develop a radiation-driven outflow.
%
%
\begin{figure*}
\centering\includegraphics[width = 1.0 \textwidth]{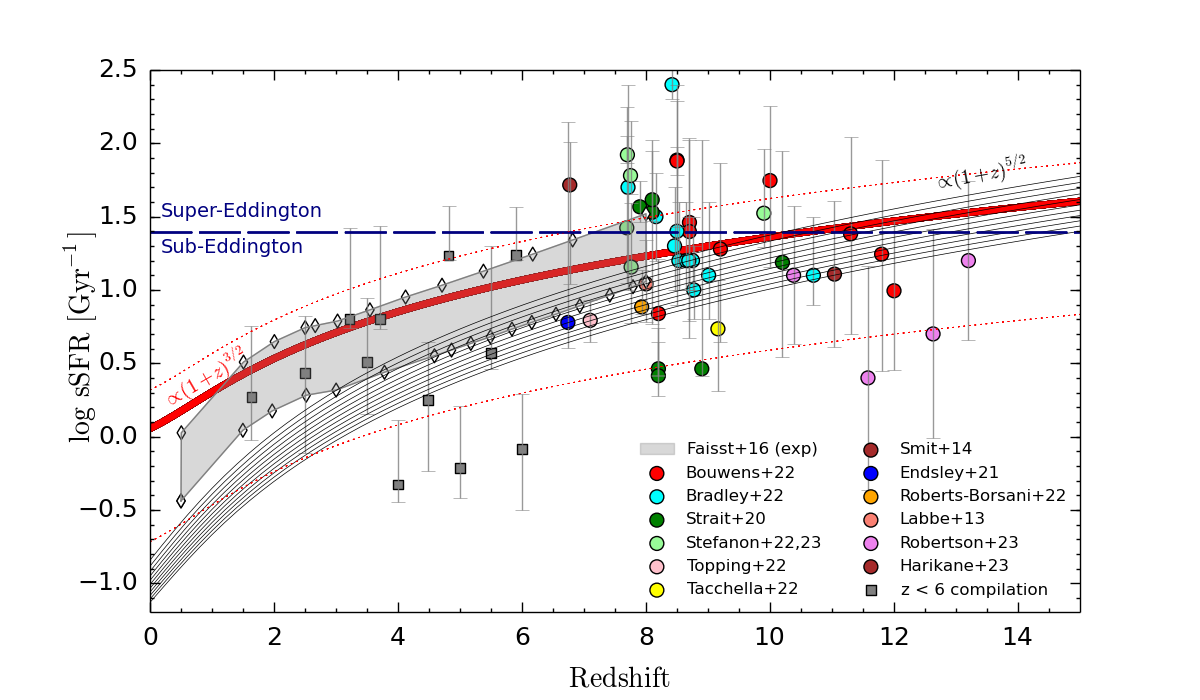}
\caption{Redshift evolution of the sSFR for the two models explored here. The red line corresponds to the fiducial model in which most stars formed over the last free-fall time (eq. \ref{eq:sSFR}) along with $\pm 1\sigma$ uncertainties (red dotted) derived from data in $8<z<10$. Black lines account for halo mass growth (eq. \ref{eq:Mhalo}) and are shown for present day halo values of $\log (M_0/\msun) = {10-15}$  (bottom to top) in 0.5 dex intervals; these curves approximately evolve as $(1+z)^{5/2}$. Models are compared with available data points for high-$z$ galaxies from \citet{Bouwens22a, bradley2022, Strait20, Stefanon22, Stefanon23, topping2022, Tacchella22, Smit14, endsley2021, Roberts22, Labbe13, Robertson23, Harikane23}. The horizontal dashed line is the critical sSFR (eq. \ref{eq:ssfr_thresh}) above which outflows develop. 
}
\label{Fig:sSFR_vs_z}
\end{figure*}

\section{Specific star formation rate}\label{sec:sSFR}
Given the result in eq. \ref{eq:ssfr_thresh}, we turn our attention to the redshift evolution of the sSFR. We follow \citet[][]{Ferrara23}, and write the mean SFR in a halo of total mass $M$ as 
\be
{\rm SFR} = \epsilon_\star f_b \frac{M}{t_{\rm ff}},
\label{eq:SFR}
\ee
where $\epsilon_\star$ is the \textit{instantaneous} star formation efficiency (i.e. at the time of observation), and $f_b = \Omega_b/\Omega_m = 0.158$. The gas free-fall time in halos, $t_{\rm ff} = (4\pi G \rho)^{-1/2}$, can be conveniently written as $t_{\rm ff}= \zeta H(z)^{-1}$, where $H(z)^{-1}$ is the Hubble time at $z$, and $\zeta=0.06$. We can then rewrite eq. \ref{eq:SFR} as
\be\label{eq:SFR1}
{\rm SFR} = 22.7 \left(\frac{\epsilon_*}{0.01}\right)\left(\frac{1+z}{8}\right)^{3/2} M_{12}\quad\msunyr.
\ee


Before we proceed we pause to test the model predictions against the $z\simeq 11.5$ UV luminosity function, and cosmic star formation rate density data. These comparisons are shown in Fig. \ref{Fig:model_test}. Note that the blue curves \textit{are not} a fit to the data, i.e. we did not vary any of the parameters of the model provided in \citet[][]{Ferrara23}. In spite of that, the predictions of the model perfectly match both data sets. On this basis, we can confidently use the model to derive the sSFR.

The galaxy stellar mass, $M_\star$, can be computed in two different ways. First, assume that star formation proceeds at a constant rate, given by eq. \ref{eq:SFR}, over the last free-fall time before the observed epoch. At $z=8-10$ this hypothesis implies a mass assembly of $\simeq 50$ Myr, which is consistent with the typical formation times estimated through non-parametric star formation histories \citep[see their Tab. 6]{Bouwens22b}. 
It follows that
\be
M_\star = {\rm SFR} \,t_{\rm ff} =  \langle\epsilon_\star\rangle f_b M,
\label{eq:Mstar}
\ee
and 
\be
{\rm sSFR} = 0.64\, \frac{\epsilon_\star}{\langle\epsilon_\star\rangle}\,(1+z)^{3/2}\, \rm Gyr^{-1},
\label{eq:sSFR}
\ee
where $\langle\epsilon_\star\rangle$ is the efficiency averaged over a free-fall time.
Note that if $\epsilon_\star \simeq \rm const. = \langle\epsilon_\star\rangle$, sSFR is independent of mass and star formation efficiency and equal to $t_{\rm ff}^{-1}$,  As the sSFR is a function of redshift only, and grows monotonically, eq. \ref{eq:ssfr_thresh} implies that \textit{on average} galaxies become super-Eddington for $z \simgt 10$ (see Fig. \ref{Fig:sSFR_vs_z}). We will refer to eq. \ref{eq:sSFR} as the \quotes{\textit{fast formation}} model. This result is essentially the same found by \citet{Tacchella18}, but the redshift evolution of the sSFR is shallower than found by \citet{Dekel13}, ${\rm sSFR} \propto (1+z)^{5/2}$.

A second way to estimate the stellar mass includes a treatment of the halo mass growth history. We start by writing
\be
M_\star(z) = \int_{z_{\star}}^z  {\rm SFR}\,\frac{dt}{dz'} dz' = \frac{\epsilon_\star f_b}{\zeta} \int_{z_{\star}}^z \frac{M(z')}{(1+z')} dz'
\label{eq:Mstar_1}
\ee
where $z_\star$ is the redshift at which the halo virial temperature reaches $T_{\rm vir} = 10^4$ K, thus enabling Ly$\alpha$ cooling\footnote{The results are almost insensitive to the value of $z_\star$}, and hence star formation. We have also used the relation $dt/dz = - [H(z)(1+z)]^{-1}$, and substituted eq. \ref{eq:SFR1} in the equation above. The halo mass growth with redshift can be obtained by time-integrating the cosmological accretion of dark+baryonic matter. A handy fit to $M(z)$ is provided by \citet{Correa15}:
\be
M(z) = M_0 (1+z)^\alpha e^{\beta z},
\label{eq:Mhalo}
\ee
where $M_0$ is the present-day halo mass, and $(\alpha, \beta)$ are parameters that depend on $M_0$, cosmology and the linear matter power spectrum. These are provided\footnote{For reference, when averaged over $8 < \log (M_0/\msun) < 14$, $(\alpha, \beta)=(0.25, -0.75)$.} in Appendix C of \citet{Correa15}. By combining eq. \ref{eq:SFR} and \ref{eq:Mstar_1} we obtain the evolution of the sSFR shown in Fig. \ref{Fig:sSFR_vs_z}. We will refer to such result as the \quotes{\textit{slow formation}} model. In this case, sSFR depends on halo mass as well as on redshift. In addition, with respect to the previous scenario (eq. \ref{eq:sSFR}), accounting for halo growth produces a steeper redshift dependence, approximately $\propto (1+z)^{5/2}$.  

In spite of these differences, both models predict that on average high-$z$ galaxies become super-Eddington. In fact, the \textit{slow formation} model predicts that halos with $M_0=10^{10} \msun$ ($M_0=10^{15} \msun$) cross the $\rm sSFR = 25 \,\rm Gyr^{-1}$ threshold at $z=14.5\, (10)$. 

At this stage the available data cannot uniquely discriminate between the two models shown in Fig. \ref{Fig:sSFR_vs_z}. Arguments in favor of a flatter $(1+z)^{3/2}$ evolution have nevertheless been brought by a number of works. \citet{Faisst16} find that sSFR $\propto (1+z)^{3/2}$ at $z >2.2$ consistent with a fast mass build-up in high-$z$ galaxies within e-folding times of $\approx 100$ Myr, consistent with our eq. \ref{eq:sSFR}. A very similar flatter trend, ${\rm sSFR} \propto (1+z)^{1.6}\, \rm Gyr^{-1}$, has been obtained by \citet{Tacchella18} and \citet{Hsiao22}. As noted, the different behavior of the models has little impact on the main point here, i.e. that a large fraction of $z \simgt 10$ galaxies should develop an outflow.  Hence, in the following we will concentrate on the \textit{fast formation} scenario.  

%
%
\begin{figure}
\centering\includegraphics[width = 0.45 \textwidth]{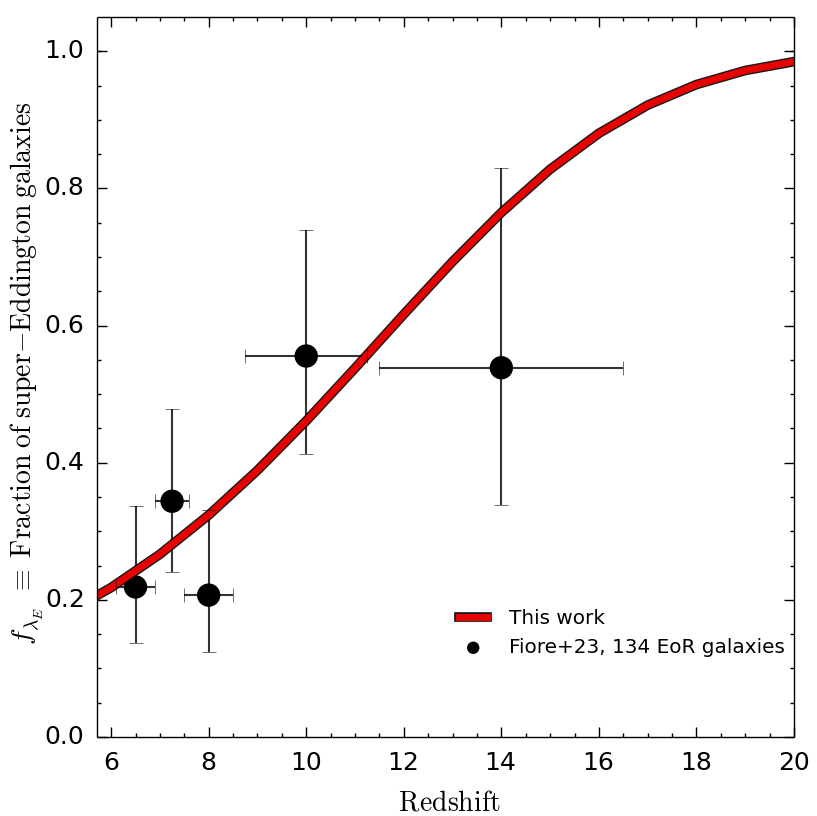}
\caption{Predicted redshift evolution of the fraction of super-Eddington galaxies (i.e. $\lambda_E^{\rm eff}>1$) in which radiation-driven outflows are expected in the fast formation model (eq. \ref{eq:sSFR}, red curve), compared with available data for 134 galaxies at $z>6.5$ \citep{Fiore23}.      
}
\label{Fig:Super_Edd_frac}
\end{figure}

\subsection{Specific star formation rate scatter}
Eq. \ref{eq:sSFR} provides a very good match to the mean sSFR trend, but it does not account for the observed sSFR scatter at fixed redshift. The scatter naturally arises in the model if the efficiency varies over $t_{\rm ff}$, i.e. $\langle \epsilon_\star \rangle \neq \epsilon_\star$. Stated differently, while the SFR depends on the instantaneous star formation efficiency, the stellar mass is sensitive to its past variations. The latter are likely regulated by poorly understood feedback processes, causing the system to experience a stochastic star formation history \citep{Pallottini23, Shen23, Mirocha23}, or even multiple subsequent phases of quiescent and active star formation \citep{Gelli23, Gelli23b, Kobayashi23}.  

For example, \citet{Looser23} reported the discovery of the quiescent low-mass galaxy JADES-GS-z7-01-QU at redshift $z=7.3$ (see also \citealt{Carnall23, Strait23} for similar systems). \citet[][]{Gelli23}  using zoom-in simulations (see also \citealt{Dome23}) showed that the fraction of time spent in an active phase increases with the stellar mass from $f_{duty}\approx 0.6$ at $M_\star\approx 10^{7.5}\msun$ to $\approx 0.99$ at $M_\star\geq 10^{9}\msun$, and it is in agreement with the value $f_{duty}\approx 0.75$ estimated for JADES-GS-z7-01-QU.

Rather than attempting to model such complex situation we resort to an empirical approach, i.e. to calculate the scatter of the model from the observed sSFR of galaxies in the relevant range $8 < z < 10$. We use a reduced chi-square to determine the variance of the observed sSFR from
the predicted one eq. \ref{eq:sSFR}: 
\be
\sigma^2 = \frac{1}{N-1}\sum_i \frac{({\rm sSFR} - {\rm sSFR}_i)^2}{\sigma_i^2}, 
\label{eq: variance}
\ee
where sSFR$_i$ and $\sigma_i$ are the $N=23$ observational points and errors shown in Fig. \ref{Fig:sSFR_vs_z}, respectively. We find that the fractional standard deviation is 83\%, and add it to eq. \ref{eq:sSFR} assuming it to be redshift-independent (red dotted lines in Fig. \ref{Fig:sSFR_vs_z}).   

Assuming a normal distribution of sSFR  with standard deviation $\sigma$, we can compute the fraction of galaxies, $f_{\lambda_E}$, at any given redshift for which $\lambda_E^{\rm eff}>1$. This is shown in Fig. \ref{Fig:Super_Edd_frac}, where it is also compared with the observational data collected in \citet{Fiore23}. We see that $f_{\lambda_E}$ increases with redshift following the overall increasing sSFR trend of the mean. At $z=6$ we find $f_{\lambda_E}=0.2$; at $z=10 \, (14)$ $f_{\lambda_E}$ reaches 45\% (76\%). 

Thus, according to our model, radiation-driven outflows should be very common in EoR galaxies, and become prevalent in galaxies at $z\simgt 10$. Their presence is solely determined by a high sSFR, and is independent of the galaxy mass. Note, however, that super-Eddington conditions might be found, albeit much more rarely, also at lower redshifts, provided that $\rm sSFR > sSFR^\star$.      

\section{Radiation-driven outflows}\label{Sec:outflows}
We now derive the properties of the radiation-driven outflows from super-Eddington galaxies. We suppose that the radiation from the stars acts on a geometrically thin, but initially UV- or even IR-optically thick dusty gas shell of mass $M_{\rm sh}$, thus transferring its momentum to both dust and gas. The shell dynamics is governed by the momentum equation:
\begin{equation}
M_{\rm sh} v \frac{dv}{dr} = f(\tau) \frac{L_{\rm bol}}{c} - G \frac{M_\star M_{\rm sh}}{r^2},
\label{eq:momentum}
\end{equation}
where $v$ is the shell velocity, and 
\be
f(\tau) = (1-e^{-\tau_{1500}}) + \tau_{\rm IR}
\label{eq:ftau}
\ee
is a function of the UV (1500 \AA) and IR optical depths \citep{thompson2015, Ishibashi2018, Costa18}. The first term in eq. \ref{eq:ftau} accounts for single-scattering momentum transfer; the second one quantifies the momentum exchange between trapped IR radiation and the outflow. 

%
%
\begin{figure}
\centering\includegraphics[width = 0.45 \textwidth]{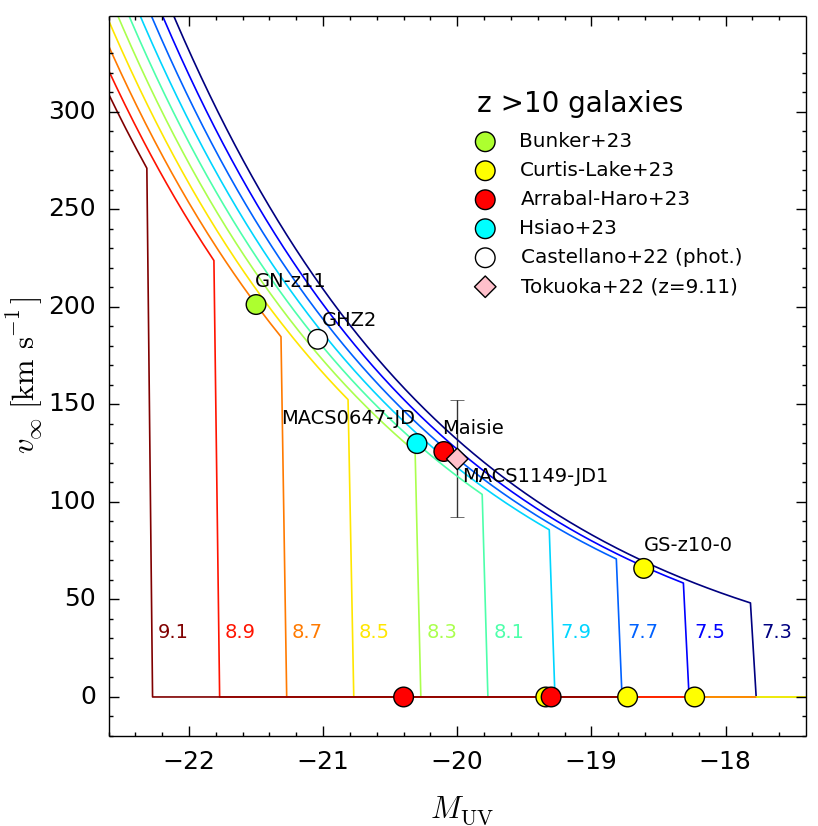}
\caption{Outflow terminal velocity (eq. \ref{eq:terminal_v1}, assuming $\langle \epsilon_\star \rangle = 0.015, f_M=0.18$, as appropriate for GN-z11) vs. UV magnitude for different values (colored numbers) of the log of the stellar mass in solar units; $v_\infty = 0$ denotes the sub-Eddington regime in which no radiation-driven outflows are expected. Circles are the predicted values of $v_\infty$ for the 9 spectroscopically confirmed galaxies presently available at $z>10$; four of them, GN-z11, MACS0647-JD, Maisie, GN-z10-0, are predicted to show outflow signatures, along with the photometric candidate GHZ2. Data are taken from \citet{Curtis23, Bunker23, Arrabal23, Harikane23, Hsiao23, Castellano22}. Shown is also the lensed ($\mu=4.7$, \citealt{Tokuoka22}) galaxy MACS1149-JD1 (\citealt{Stiavelli23}, $z=9.11$) specifically discussed in Sec. \ref{sec:discussion}. }
\label{Fig:Terminal_vel}
\end{figure}

Adopting a Milky Way extinction curve with a dust-to-gas ratio $D_{\rm MW}=1/162$, the dust mass absorption coefficient at 1500 \AA\, is $\kappa_{1500} = 1.26\times 10^5\ {\rm cm}^2 {\rm g}^{-1}$ \citep{Ferrara22a}. The analogous IR coefficient, $\kappa_{\rm IR}(\nu) = \kappa_{158}(\nu/\nu_{158})^{\beta_d}$, is pivoted at wavelength $\lambda_{158} = c/\nu_{158} =158\mu$m. For a MW curve $\kappa_{158}=10.41\, {\rm cm}^2 {\rm g}^{-1}$, and $\beta_d=2.03$ \citep{Weingartner01}.
As dust in early galaxies is found to be increasingly warm at high-$z$ reaching values of $50-70$ K at $z \simeq 10$ \citep[][and references therein]{Sommovigo22}, we adopt $T_d=60$ K as a fiducial value, and compute $\kappa_{\rm IR}$ at the peak wavelength of the grey-body spectrum, $\lambda_p = 0.29/T_d = 48\, \mum$. This yields $\kappa_{\rm IR}(\nu_p) = 117\ {\rm cm}^2\, {\rm g}^{-1}$. 

As long as the galaxy is significantly super-Eddington, the gravity term in eq. \ref{eq:momentum} can be neglected with respect to $f(\tau)L_{\rm bol}/c$, and it is easy to see that the outflow reaches a terminal velocity
\be
v_\infty^2 \approx f(\tau) \frac{L_{\rm bol}}{c} \frac{r_{\tau=1}}{M_{\rm sh}},
\label{eq:terminal_v}
\ee
where $r_{\tau=1}$ is the radius at which the outflow becomes optically thin due to expansion, i.e. $\max(\tau_{1500},\tau_{\rm IR}) = \tau_{1500} = 1$. Following \citet{Ziparo23}, we write 
\begin{equation}
\tau_{1500} = \frac{\kappa_{1500}\xi_d}{4\pi r_e^2}  M_\star,    
\label{eq:tau1500}
\end{equation}
where $\xi_d = 1/529$ is the dust-to-stellar mass ratio due to supernova (SN) dust production for a Salpeter $1-100\, \msun$ IMF, and assuming a dust yield/SN of 0.1 $\msun$; $r_e$ is the stellar effective radius.  Recalling that at high redshifts $r_e \approx 0.1$ kpc for a galaxy with stellar mass $M_* \approx 10^9 \msun$ \citep{Pallottini22, Adams22, Bunker23, Ono22}, eq. \ref{eq:tau1500} numerically yields 
\begin{equation}
\tau_{1500} = 396 \left(\frac{M_\star}{10^9 \msun}\right) \left(\frac{100\ {\rm pc}}{r_e}\right)^{2}.    
\label{eq:tau1500_num}
\end{equation}
The above result shows that super-early, massive galaxies are heavily obscured both in the UV, and possibly also in the IR ($\tau_{\rm IR} = \tau_{1500}/10^3$) during their initial evolutionary phases, and before the outflow is launched. From eq. \ref{eq:tau1500_num} it also follows that $r_{\tau=1} =2 (M_\star/10^9 \msun)^{1/2}\, \rm kpc$. 

To evaluate eq. \ref{eq:terminal_v}, the final ingredient is $M_{\rm sh}$. We assume that a fraction $f_M$ of the gas left over from star formation is accelerated in the expanding shell.  Hence,  $M_{\rm sh} = f_M [(1- \langle \epsilon_\star \rangle)/\langle \epsilon_\star \rangle] M_\star \equiv (f_M/g_\epsilon) M_\star$. Taking the best studied $z>10$ galaxy, GN-z11 \citep{Bunker23}, from the observed Ly$\alpha$ line profile and displacement (see Sec. \ref{sec:Lya}) we deduce $f_M=0.18$; $\langle \epsilon_\star \rangle = 0.015$ is obtained by matching\footnote{We further derive a halo mass $M \simeq 10^{11.5} \msun$ by abundance matching of the UV LF (see Fig. \ref{Fig:model_test}).} the predicted SFR of GN-z11 (eq. \ref{eq:SFR1}) with the observed one ($18.78\, \msunyr$). The final expression for the outflow terminal velocity is 
\be
v_\infty = 830\ \sqrt{\frac{g_\epsilon}{f_M}} \left(\frac{\rm sSFR}{\rm sSFR^\star}\right)^{1/2}\left(\frac{M_\star}{10^9 \msun}\right)^{5/12} \rm km\ s^{-1},
\label{eq:terminal_v1}
\ee
having further used the scaling $r_e \propto M_\star^{1/3}$ which implicitly assumes that the effective radius is proportional to the galaxy virial radius, $r_e \approx 0.01 r_{\rm vir}$ \citep{Shibuya15, Ono22}. Note that we have normalized the sSFR to the critical value, $\rm sSFR^\star = 25\ \rm Gyr^{-1}$, obtained in eq. \ref{eq:ssfr_thresh}; also, the dependence on stellar mass is very weak, $\propto M_\star^{-1/12}$.  
The terminal velocity increases with the star formation efficiency (via $g_\epsilon$), and $v_\infty \propto \langle \epsilon_\star\rangle^{1/2}$.  Outflows clear the dust and gas from the galaxy main body in a short timescale, $t_c \approx r_{\tau=1}/3 v_\infty = 3.1\ (200\, {\rm km\ s}^{-1}/v_\infty)\ \rm Myr$, thus making the galaxy UV spectrum very blue, as preliminary found by \citet{Cullen23} and \citet{Topping23}. 

Fig. \ref{Fig:Terminal_vel} shows $v_\infty$, derived from eq. \ref{eq:terminal_v1}, as a function of the UV magnitude of galaxies with stellar masses in the range $10^{7.3-9.1}\, \msun$. Generally, the outflow velocity increases towards brighter magnitudes with a moderate scatter introduced by the stellar mass. Outflows can occur at any $M_{\rm UV}$ but more massive galaxies do not produce an outflow ($v_\infty = 0$) until they reach bright magnitudes. 

For example, the outflow of a galaxy like GN-z11 ($M_{\rm UV} = -21.5$, $M_\star = 10^{8.7}\ \msun$) is predicted to reach $v_\infty \simeq 200 \, \kms$. Note that $v_\infty$ is smaller than the circular velocity of GN-z11 halo, $v_c = 324\, \rm km\ s^{-1}$, having assumed a halo mass\footnote{Our estimate is based on the halo abundance matching of the LF \citep{Ferrara23}. Given the measured stellar mass, and the extrapolated stellar-halo mass relation in \citet{Behroozi13}, \citet{Scholtz23} find instead $M=2.96^{+0.44}_{-0.49}\times 10^{10}\msun$.} $M \simeq 10^{11.5} \msun$. This indicates that dust and gas are efficiently removed from the galaxy by the outflow, but they are likely to remain bound to the system as a spatially extended component or a fountain-like flow.
This prediction is also consistent with the results by \citet[][Fig. 7]{Xu23}, who find that in their sample of 30 outflowing galaxies at $3 < z < 9$, almost all of the outflows cannot escape from the halo gravitational potential.

In Fig. \ref{Fig:Terminal_vel} we consider the 9 spectroscopically confirmed galaxies (see Tab. \ref{Tab:properties} for a summary of their properties) currently available at $z>10$, plus an interesting photometric candidate, GHZ2 \citep[][$z=12.2$]{Castellano22}. We find that 5 out of these galaxies (i.e. 50\%) have super-Eddington luminosities, in excellent agreement with the $z\simeq 10-12$ model predictions shown in Fig. \ref{Fig:Super_Edd_frac}. Their outflow velocities largely depend on galaxy sSFR, and range from $65\, \kms$ for GS-z10-0, to $200\, \kms$ for GN-z11. The remaining five galaxies are instead in the sub-Eddington (no outflow) regime. 

\subsection{Testing the outflow model}
The predicted presence of radiation-driven outflows in high-$z$ galaxies can be confronted with observations. Luckily, and thanks to the advent of {\it JWST}, high-quality spectroscopic data have recently become available. \citet[][see also \citealt{Zhang23, Xu23}]{Carniani23} have investigated the ionized gas outflows in 52 galaxies, part of the JADES sample, at $3 < z < 9$ with stellar mass $10^{7-9}\ \msun$. They detected broad line components of optical nebular lines, which they interpret as an outflow signature, in $\approx 25-40\%$ of the systems. This value represents a lower limit to the incidence if outflows are anisotropic. 

\begin{figure}[t]
\includegraphics[width = 0.48 \textwidth]{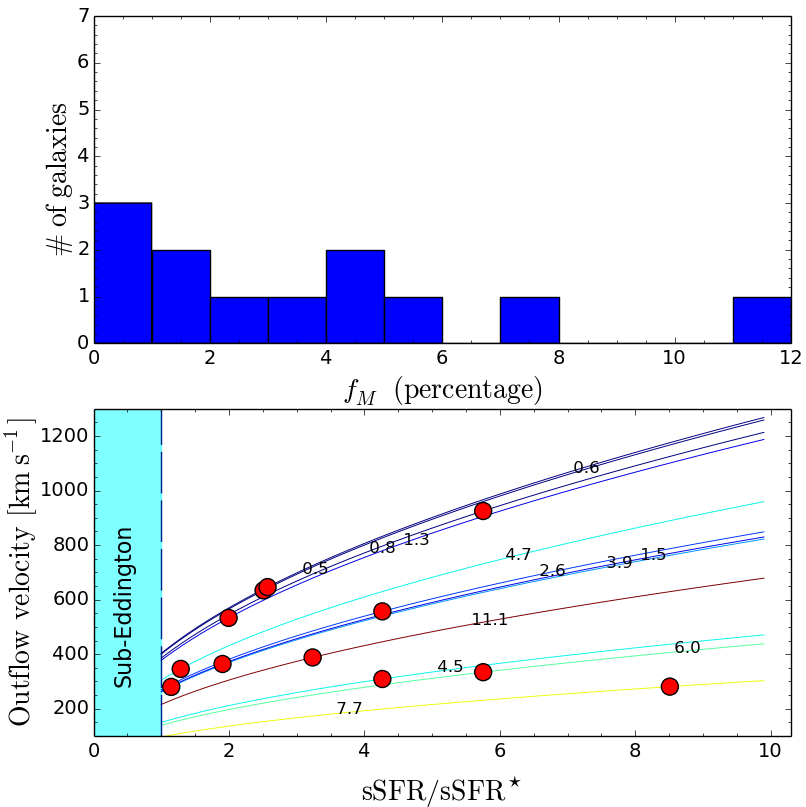}
\caption{\textit{Bottom panel}: Model predictions for the outflow terminal velocity (eq. \ref{eq:terminal_v1}, lines) as a function of the sSFR normalized to the critical value sSFR$^\star$ (eq. \ref{eq:ssfr_thresh}). Galaxies for which sSFR/sSFR$^\star>1$ are in the super-Eddington regime; also shown (cyan band) is the sub-Eddington region. The predictions are compared to {\it JWST}/JADES data \citep[red points]{Carniani23} for 12 galaxies in $3 < z < 9$ with measured outflow velocity. The data are also used to derive the fraction of halo gas entrained by the outflow, $f_M$ (see eq. \ref{eq:terminal_v1}), whose (percentage) value is shown by the number on each curve. \textit{Top}: Histogram of the $f_M$ values derived in the bottom panel.   
}
\label{Fig:carniani}
\end{figure}

Interestingly, the 12 galaxies in the sample that are super-Eddington (sSFR/sSFR$^\star = 1-8.5$) all show the presence of an outflow with velocities in the range  $280-925\, \kms$ (Fig. \ref{Fig:carniani}). This strongly suggests that radiation pressure from young stars in the galaxy drives the outflow. The observed velocities may be used to constrain the fraction of \textit{halo gas} entrained by the outflow, $f_M$, which is a free-parameter in eq. \ref{eq:terminal_v1}. We find that $f_M$ is typically $1-10$\%, with an average value of 4\%; its distribution is shown in the top panel of Fig. \ref{Fig:carniani}. This value roughly corresponds to the entire galaxy gas mass.
%
%
%
%

%
%
\begin{figure*}[t]
\begin{tabular}{lr}
\includegraphics[width = 0.48 \textwidth]{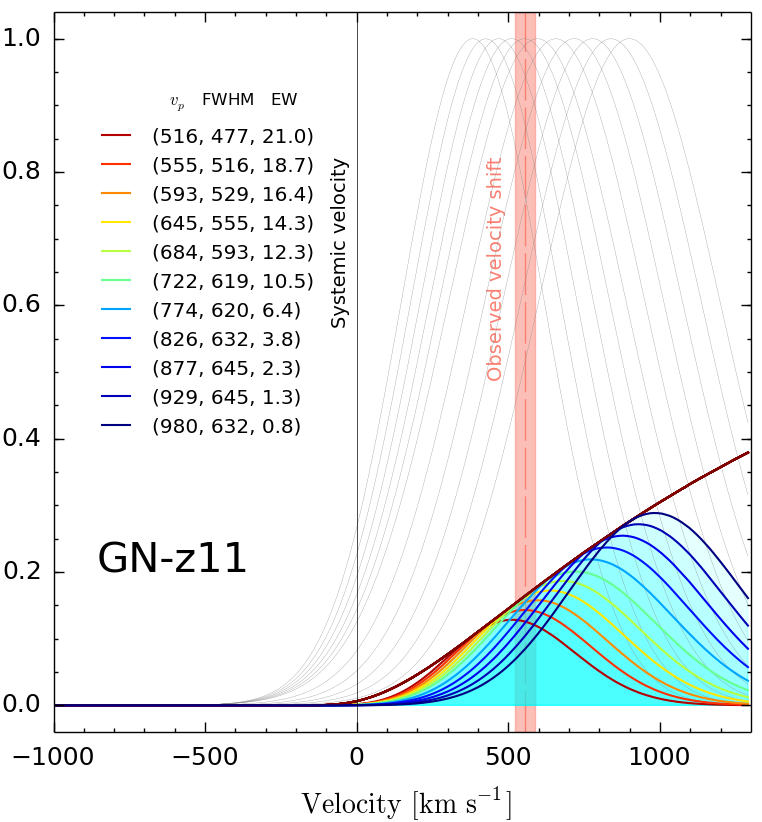}
\includegraphics[width = 0.48 \textwidth]{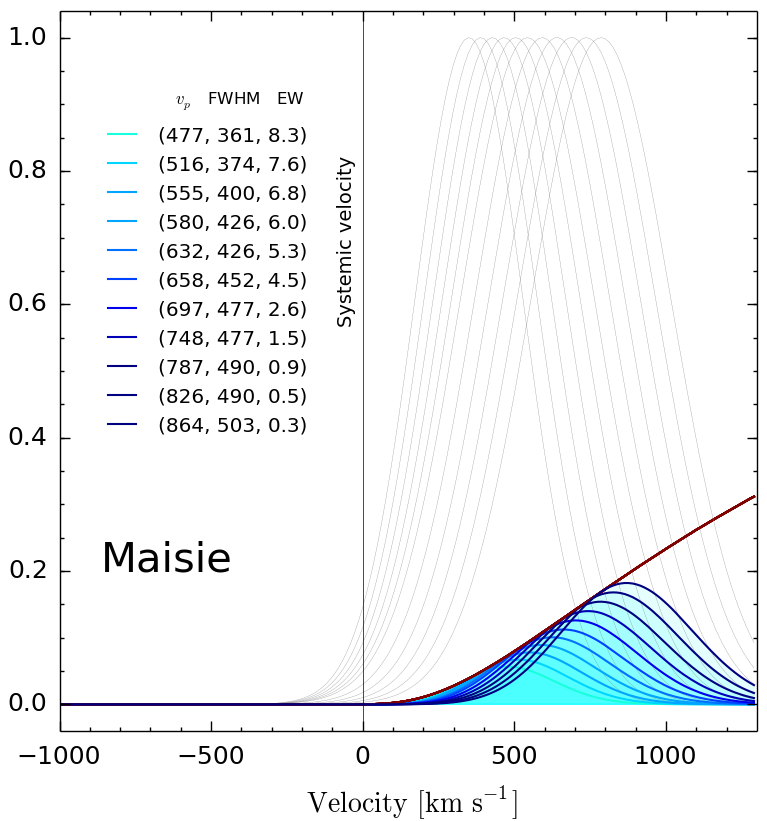}
\end{tabular}
\caption{Predicted profiles of the observed Ly$\alpha$ line from two super-Eddington galaxies GN-z11 \citep{Bunker23} at $z=10.6$, and Maisie \citep{finkelstein2022, Arrabal23} at $z=11.4$. In each panel, different curves from left to right refer to ISM neutral hydrogen column densities $N_{\rm HI} = 10^{20+0.1j} \rm cm^{-2}$, with $j=0,1,..,10$. The grey lines show the Ly$\alpha$ profile emerging from the galaxy, also accounting for the outflow velocity; colored lines represent the observed profile after transmission through the IGM whose GP damping wing including the surrounding HII bubble, $e^{-\tau_{\rm GP}}$, is shown by the red curve. For each observed curve we show the values of the line peak velocity shift, $v_p$, FWHM  (both in $\rm [km\, s^{-1}]$), and EW in [\AA]. For GN-z11 the observed values \citep{Bunker23, Scholtz23} are ($v_p$, FWHM, EW) = ($555 \pm 32\, \kms, 530\pm 65\, \kms, 18\pm 2$ \AA) which are very well reproduced by the curve with $N_{\rm HI} = 10^{20.1} \rm cm^{-2}$. At the same $N_{\rm HI}$, we predict that the Ly$\alpha$ EW for Maisie is 7.6 \AA. 
}
\label{Fig:Lya_profiles}
\end{figure*}

\section{Lyman Alpha visibility in the EoR}\label{sec:Lya}
We showed that radiation-driven outflows are common among super-early galaxies. If so, they can dramatically enhance the transmission of the Ly$\alpha$ line emitted by these sources even if the intergalactic medium (IGM) is almost neutral, as expected at $z\simgt 10$. This fact was already noted by, e.g. \citet{Dijkstra10}, who pointed out that scattering off outflows causes the Ly$\alpha$ flux to emerge from galaxies at frequencies where the Gunn–Peterson (GP) optical depth is highly reduced. 

To quantify the Ly$\alpha$ visibility, we concentrate on two super-early galaxies, GN-z11 \citep[$z=10.6$]{Bunker23}, and Maisie \citep[$z=11.4$]{Finkelstein22, Arrabal23} for which exquisite spectroscopic data are available. The results are displayed in Fig. \ref{Fig:Lya_profiles}. Moreover, we have considered also the other three $z>10$ super-Eddington galaxies (GHZ2, MACS0647-JD, GS-z10-0) included in Fig. \ref{Fig:Terminal_vel}. The results for these five galaxies are summarized in Tab. \ref{Tab:properties}.  

To derive the observed Ly$\alpha$ line profile for each galaxy three steps are required: (a) compute the outflow velocity, $v_\infty$. This step has been performed in eq. \ref{eq:terminal_v1}; (b) compute the line profile emerging from the galaxy after processing by radiative transfer effects (\HI scattering, dust absorption) in the outflow; (c) convolve the emergent line profile with the IGM attenuation due to the GP optical depth $\tau_{\rm GP}$ to determine the observed profile. We next describe how steps (b) and (c) are performed.

\subsection{Emergent Ly$\alpha$ profile}
To compute the emergent Ly$\alpha$ profile we use the results by \citet{Orsi12}. They used a Monte Carlo Ly$\alpha$ radiative transfer code to determine the line properties depending on the outflow velocity, $v_\infty$, and \HI column density, $N_{\rm HI}$. As a function of these two quantities\footnote{The outflow geometry is modelled either as a wind or a shell; the differences between the two cases are small, and we adopt here the shell geometry case.}, they provide the emergent Ly$\alpha$ line median velocity shift with respect to systemic ($\tilde v_p^{\rm ISM}$), full-width half-maximum (FWHM$^{\rm ISM}$), and Ly$\alpha$ escape fraction, ($f_\alpha^{\rm ISM}$), i.e the fraction of the line luminosity that escapes the system and into the IGM.  We linearly interpolate the results given in \citet[][Fig. 2 for $\tilde v_p^{\rm ISM}$, FWHM$^{\rm ISM}$; Fig. 3 for $f_\alpha^{\rm ISM}$]{Orsi12}. 

Moreover, Ly$\alpha$ can also be extincted by dust, whose abundance is approximately $\propto Z$. \citet{Orsi12} considered metallicities in the range $Z=0.5-1\, \zsun$; however, they note that this choice only affects the results at HI column densities, $N_{\rm HI} \simgt 10^{21} \rm cm^{-2}$, that are not relevant for our results below. We assume that the emergent profile is a gaussian\footnote{In principle, \citet{Orsi12} also give the level of asymmetry of the line. However, the relatively poor sampling and the small values obtained prevent us from robustly including  this information in the model.} centered (in velocity space) at $\tilde v_p^{\rm ISM}$ and full width equal to FWHM$^{\rm ISM}$. Finally, we use $f_\alpha^{\rm ISM}$ to compute the line EW (see below).  

%
%
\begin{table*}
\begin{minipage}{170mm}
\begin{center}
\caption{Observed and derived properties of currently available $z>10$ spectroscopically confirmed super-Eddington galaxies, also including the photometric candidate GHZ2.}
\begin{tabular}{cccccccccccccc}
\hline\hline
\multicolumn{6}{c}{\textit{Measured}}& \multicolumn{1}{c}{ID\#} & \multicolumn{7}{c}{\textit{Derived}}\\
\cline{1-6} \cline{8-14} 
$z$ & $\beta$& $\log M_\star$&   SFR    &   sSFR    &$\log \xi_{\rm ion}$&      &$v_\infty$&$f_{\rm esc}$&$R_{b}$&$v_p$ &FWHM &EW &$f_\alpha$\\
\hline 
    &        &   $\msun$            & $\msunyr$& Gyr$^{-1}$&           erg$^{-1}$ Hz         &      & $\kms$   &             & pMpc  &$\kms$&$\kms$&\AA&          \\
\hline
10.60& $-$2.40 &  8.70          &   15.7 &  31.3     &        25.70       &GN-z11&  200     &   0.03      & 0.21  & 555  & 516 &18.7& 0.069    \\
11.40& $-$2.32 &  7.78          &    4.2 &  69.7     &      25.70$^{(a)}$ &Maisie&  124     &   0.02      & 0.11  & 516  & 384 & 7.6& 0.028    \\
10.38& $-$2.49 &  7.58          &    1.1 &  28.9     &      25.46       &GS-z10-0&   65     &   0.04      & 0.08  & 490  & 335 & 4.0& 0.026    \\
10.17& $-$2.60 &  8.10          &    5.1 &  40.5     &      25.20    &MACS0647-JD&  129     &   0.05      & 0.13  & 503  & 387 & 3.4& 0.040    \\
12.20& $-$3.00 &  8.08          &   10.1 &  84.0     &      25.70$^{(a)}$   &GHZ2&  182     &   0.16      & 0.26  & 529  & 503 &17.6& 0.065    \\
\hline
\label{Tab:properties}
\end{tabular}
\end{center}
\end{minipage}
\tablecomments{Columns (1)-(6) contain the observed properties of the galaxies taken from \citet{Bunker23,Arrabal23, Curtis23, Hsiao22, Hsiao23, Castellano22, Harikane23}; Columns (8)-(10) give the outflow terminal velocity, LyC escape fraction, and \HII bubble radius, respectively; Columns (11)-(14) give the model predictions for the observed Ly$\alpha$ line peak velocity shift, FWHM, EW and escape fraction, respectively; they are computed for the same outflow \HI column density, $\log N_{\rm HI}=21.1$, fitting the observed GN-z11 line profile for each galaxy. $^{(a)}$ As for Maisie and GHZ2 the value of $\log \xi_{\rm ion}$ is unknown, we adopted the one of GN-z11.} 
\end{table*}

\subsection{Observed Ly$\alpha$ profile}
The emergent line profile is further processed by the intervening neutral IGM along the line of sight, also considering the fact that LyC photons escaping from GN-z11 carve an approximately spherical ionized bubble around the galaxy\footnote{Other fainter companion galaxies residing in the bubble might also contribute ionizing photons  \citep{Castellano16, Castellano22}. We neglect this effect here. Therefore $R_b$ should be considered as a lower limit to the actual bubble size.}. The bubble radius might be computed from the observed properties of GN-z11 as follows:
\begin{equation}
R_b = \left(\frac{3 N_{\rm ion} f_{\rm esc} t_s}{4 \pi n_e}\right)^{1/3},
\label{eq:Rb}
\end{equation}
where $N_{\rm ion} = \xi_{\rm ion} L_{1500}$ is the LyC photon emission rate.
The LyC escape fraction, $f_{\rm esc}$, is obtained from the UV spectral slope $\beta$ using the \citet{Chisholm22} relation: $\log f_{\rm esc} = -4.44 - 1.22 \beta$. We have slightly corrected the relation to match the GN-z11 value $f_{\rm esc}=0.03$ deduced from the H$\beta$ equivalent width \citep{Bunker23}. The timescale, $t_s$, is taken as the minimum between the stellar age and the recombination time,
\begin{equation}
t_s =  \min(t_\star, 1/\alpha_B n_e),
\label{eq:ts}    
\end{equation}
where $t_\star \approx t_{\rm ff}$ (see Sec. \ref{sec:sSFR}), $\alpha_B = 2.6\times 10^{-13} (T/10^4\, \rm K)^{-3/4}\, \rm cm^3\, s^{-1}$ is the Case-B recombination coefficient, and $n_e \approx n_{\rm H}$ is the mean (electron) density of the IGM at the redshift of the emitter. Eq. \ref{eq:ts} accounts for the fact that the stellar population in a galaxy might be too young to fill the galaxy Str\"omgren sphere. The final $R_b$ values are given in Tab. \ref{Tab:properties}; they are in the range $0.08-0.26$ pMpc. 

The GP scattering optical depth at the observed wavelength $\lambda_{\rm obs} > \lambda_\alpha (1+z_{\rm em})$, where $\lambda_\alpha = 1215.67$ \AA\, is the restframe Ly$\alpha$ wavelength, and $z_{\rm em}$ the redshift of the source, is 
\begin{equation}
\tau_{\rm GP}(\lambda_{\rm obs}) = \int_{z_{\rm rei}}^{z_{\rm b}} dz \frac{d\ell}{dz} n_{\rm HI}(z) \sigma_\alpha\Big[\nu=\frac{c (1+z)}{\lambda_\alpha} \Big],
\label{eq:tauGP}    
\end{equation}
where $\sigma_\alpha(\nu, T)$ is the Ly$\alpha$ cross section assuming an IGM temperature $T=100$ K at $z\approx 10$. We assume that inside $R_b$ the IGM is fully ionized, i.e. we neglect the residual \HI in the bubble, and write the proper cosmological line element as ${d\ell}/{dz} = c[(1+z)H(z)]^{-1}$. The reionization redshift is fixed to $z_{\rm rei}=6$, and $z_b$ is defined as the redshift corresponding to the \HII bubble radius of each source. Finally, $n_{\rm HI}(z)=x_{\rm HI} n_{\rm H}$ is the \HI density at redshift $z$; we set $x_{\rm HI}=1$, consistent with recent findings at $z \simeq 10$ \citep{Bruton23}.

The observed Ly$\alpha$ profile is then computed by multiplying the emergent profile by $e^{-\tau_{\rm GP}}$, i.e. the so-called GP damping wing. We also compute the fraction, $f_\alpha^{\rm IGM}$, of the emergent Ly$\alpha$ luminosity that escapes after attenuation by the IGM.   The emergent and observed Ly$\alpha$ profiles of GN-z11 and Maisie are shown as an example in Fig. \ref{Fig:Lya_profiles}, along with the corresponding GP damping wings, as a function of velocity. 

\subsection{Ly$\alpha$ emission from super-Eddington galaxies}
From the predicted Ly$\alpha$ profiles of the five super-Eddington galaxies, we can extract four quantities that can be readily compared with observations. These are the Ly$\alpha$ line peak velocity shift, $v_p$, FWHM, EW, and total escape fraction, $f_\alpha = f_\alpha^{\rm ISM} \times f_\alpha^{\rm IGM}$. 

\subsubsection{GN-z11}
Let us first concentrate on GN-z11 for which the Ly$\alpha$ has been detected by \citet{Bunker23} (Fig. \ref{Fig:Lya_profiles}, left panel). For the predicted outflow velocity, $v_\infty = 200\, \kms$, the line profile depends on the outflow \HI column density which we allow to span the range $\log N_{\rm HI} = 20-21$. Both the line peak velocity shift and full-width half-maximum increase with $N_{\rm HI}$, whereas the EW decreases due to the increased attenuation. The observed values ($v_p$, FWHM, EW) = ($555 \pm 32\, \kms, 530\pm 65\, \kms, 18\pm 2$ \AA) are simultaneously well recovered if $\log N_{\rm HI} \simeq 20.1$ (see Tab. \ref{Tab:properties} and Fig. \ref{Fig:Lya_profiles}). The predicted total Ly$\alpha$ escape fraction is $f_\alpha = 0.069$. We have checked that artificially setting $v_\infty=0$ (i.e. no outflow) it is impossible to recover the observed line parameters simultaneously. Interestingly, the derived value of $N_{\rm HI}$ is fully consistent with the one measured by \citet{Umeda23} in 26 {\it JWST}-detected galaxies at $7 < z < 12$.   

From the fit to the GN-z11 Ly$\alpha$ line profile it is found that the outflow shell contains a fraction $f_M=0.18$ of the baryonic mass of the halo. The total hydrogen column density of the shell measured at the optically thin radius, $r_{\tau=1}$, is $\log N_{\rm H} = 22.5$. By comparing with the value $\log N_{\rm HI}=20.1$ found from the Ly$\alpha$ analysis, we conclude that the outflow is largely ionized, with a neutral fraction $x_{\rm HI} = N_{\rm HI}/N_{\rm H} = 4.2\times 10^{-3}$. Such ionized state can result from either photoionization, shocks, or a combination of the two. Radiative hydro-dynamical simulations are required to clarify this issue.

The excellent match provided by the model to the observed line properties can be considered as a successful test for the radiation-driven outflows proposed here. It also points toward the importance of outflows in determining the Ly$\alpha$ visibility from $z>10$, where in principle the line transmission should be almost completely suppressed by the neutral IGM. 

\subsubsection{Preditions for Maisie and other galaxies}
Beyond explaining the GN-z11 data, our model makes detailed predictions (summarized in Tab. \ref{Tab:properties}) on the outflow-boosted Ly$\alpha$ emission from the other four super-Eddington galaxies. We start by analyzing Maisie, whose predicted Ly$\alpha$ profile is shown in Fig. \ref{Fig:Lya_profiles} (right panel). As for Maisie the Ly$\alpha$ line could have not been detected in the {\it JWST}/NIRSpec prism spectra, we cannot put constraints on the \HI column density. Rather, the values reported in Fig. \ref{Fig:Lya_profiles} and Tab. \ref{Tab:properties} can be considered as predictions of the model. For comparison sake, though, we can consider the curve with $\log N_{\rm HI}=20.1$, appropriate for GN-z11. Maisie shows a slightly lower line peak velocity shift of 516 $\kms$, and significantly narrower FWHM (384 $\kms$). As a result the line EW is smaller (7.6 \AA), with only 2.8\% of the Ly$\alpha$ photons transmitted.  

The other three super-Eddington galaxies do not differ considerably in terms of $v_p$, FWHM, and EW from Maisie, with the exception of GHZ2 (photometric redshift $z=12.4$), which instead shows properties very similar to GN-z11. Notably, its large EW = 17.6 \AA clearly illustrates the boost that outflows can provide to the line visibility. GHZ2 is predicted to have an outflow travelling at $v_\infty = 182\, \kms$, and aided by its blue colors ($\beta=-3.0$), and consequently large LyC escape fraction $f_{\rm esc}=0.16$, it manages to transmit 6.5\% of its Ly$\alpha$ photons in spite of being located at the highest redshift in the sample. 

\section{Discussion}\label{sec:discussion}
Although the radiation-driven outflow scenario discussed here provides a viable interpretation of {\it JWST} data that imply (a) an unexpected excess of bright galaxies, (b) with very blue colors, (c) some of which are able to transmit their Ly$\alpha$ line through the largely neutral IGM at $z>10$, several aspects of the model need further clarification.  

\subsection{Outflow physics}
We have shown that for dusty outflows the Eddington luminosity is reduced by a factor $A=100-600$, making radiation pressure extremely efficient. The super-Eddington condition is equivalent to the requirement that ${\rm sSFR} > {\rm sSFR}^\star \simeq 25 ({100}/{A}) \,{\rm Gyr^{-1}}$. Thus, a larger value of $A$ would favour the onset of outflows from galaxies with even more modest sSFR (see \citealt{Fiore23}, Fig. 2). The value of $A$ is uncertain as it depends on the galaxy dust-to-gas ratio, interstellar radiation field, and radiative transfer effects \citep{fabian2006, Fiore23}.  Although these aspects deserve more study, our choice of $A=100$ is very conservative.

We additionally note that even if stars in super-early galaxies formed out of a metal-free gas with no dust, radiation pressure can be initially provided -- with comparable strength -- by Ly$\alpha$ photons \citep{Tomaselli21}. This mechanism has been shown to be at work in Green Pea analogs, like  Mrk 71 \citep{Komarova21}. The interesting feature is that Ly$\alpha$ radiation pressure starts to work immediately after the formation of the first massive stars, with no delay ($\simgt 3\, \rm Myr$) associated with SN explosions and dust production. Ly$\alpha$ feedback might  effectively cap the star formation efficiency \citep{Tan08, Dekel23}. 

The Eddington condition, eq. \ref{eq:ssfr_thresh}, has been derived under the implicit assumptions that (a) gravity is dominated by stars; (b) the system is spherically symmetric. The first assumption is more delicate as dark matter and gas in the halo might also contribute to the gravitational field. To get a simple estimate of these effects, we assume that dark matter follows a singular isothermal radial density profile, 
\be
M(<r) \approx \frac {v_c^2}{G} r. 
\label{eq:isothermal} 
\ee
where $v_c$ is the halo circular velocity. We further assume that baryons initially follow the same distribution scaled by the cosmological baryon fraction, $M_b = f_b M$. 

Consider the case of GN-z11, for which $M_\star=10^{8.7} \msun$ of stars are located in an effective radius  $r_e= 64$ pc, and $v_c = 324\, \kms$. The gas mass in that volume, according to eq. \ref{eq:isothermal}, is $10^{8.4} \msun < M_\star$, implying that all the gas initially within $r_e$ (and, formally, up to $\simeq 200$ pc) has been turned into stars. Hence, gas outside this region is subject to the gravitational pull of stars \textit{and} dark matter. 

Condition (a) above is satisfied as long as stars dominate the gravitational field, i.e. for $r_g < GM_\star/v_c^2 \approx 20\, {\rm pc} < r_e$. Hence, outside $r_e$ stars are subdominant, and one should replace $M_\star$ with $M_\star + M(< r)$ in the gravity term in eq. \ref{eq:terminal_v}. However, in the derivation of $v_\infty$ we neglected the gravity effects by considering only very super-Eddington system (i.e. $\rm sSFR\gg \rm sSFR^*$); in this approximation our treatment still holds\footnote{If the halo mass is instead $2.96\times 10^{10} \msun$ \citep{Scholtz23}, then $r_g \simeq 100$ pc, essentially validating our approximation.}. 

Assumption (b) can be justified with the very compact, almost point-like, size ($\simlt 100$ pc) of super-early galaxies. Clearly, departures from spherical symmetry are also possible in case of, e.g. disk formation. One can show, that for a thin stellar disk embedded in a more extended gas layer the super-Eddington condition is the same that we use here as long as gravity is dominated by stars. 

\citet{Ziparo23} have explored the case of a disk geometry assuming a self-gravitating disk. They also conclude that galaxies with high sSFR develop radiation-driven outflows. However, \citet{Ziparo23} noted that is some extreme cases the dust production rate might exceed the outflow rate, and dust accumulates faster than it is ejected. This implies that a fraction of super-early galaxies might remain enshrouded in dust for several Myr. If spectroscopically confirmed, the sample of galaxies studied by \citet[][see also \citealt{Dressler23}]{rodighiero2022} might be a notable example of this population. 

Outflows with velocities of $\approx 800\, \kms$ have been identified in GN-z11 \citep{Maiolino23} from the analysis of CIV through. This velocity is  compatible with our predictions in eq. \ref{eq:terminal_v1} as long as $f_M$ is low, although an additional momentum injection from an alleged AGN is possible. A low $f_M$ might also indicate that the outflow is anisotropic, and able to clear dust and gas only in a channel oriented along the line of sight to the galaxy. 

As a final remark, it is also worth recalling that the predicted outflow velocities in many cases are lower than the circular velocity of the halo, implying that the dust and gas are not ejected beyond the virial radius. This is not in contrast with our argument as blue spectral colors can be obtained by pushing the dust beyond the radius $r_{\tau=1} \approx 2-3\, \rm kpc$ (eq. \ref{eq:tau1500_num}) at which it becomes optically thin. 

Once the dust distribution is effectively removed from the main body of the galaxy it will become hard to detect it from its rest-frame UV/optical extinction and/or thermal emission. A recent example can quantify this statement. \citet{Stiavelli23} have obtained {\it JWST}/NIRSpec data on the $z=9.11$ galaxy MACS1149-JD1 (JD1), and determined its dust content from the Balmer decrement. In spite of a relatively large stellar mass, $M_\star \approx 10^8\, \msun$, JD1 is consistent with having no dust. The absence of significant amounts of dust is also supported by the non-detection of FIR continuum for this object \citep{Hashimoto18, Tokuoka22}. Interestingly, the sSFR$=36.9^{+0.26}_{-0.26}\, \rm Gyr^{-1} > sSFR^{\star}$ of JD1 indicates that dust has been evacuated by a radiation-driven outflow. We also note that if the measured velocity difference between the blue- and red-shifted components
of the [OIII] 88 $\mum$ line ($122 \pm 30 \kms$, \citealt{Tokuoka22}) is interpreted as an outflow, such velocity is consistent with the predictions of Fig. \ref{Fig:Terminal_vel}, given the (de-lensed) magnitude of JD1.


%
%
\begin{figure}
\includegraphics[width = 0.48 \textwidth]{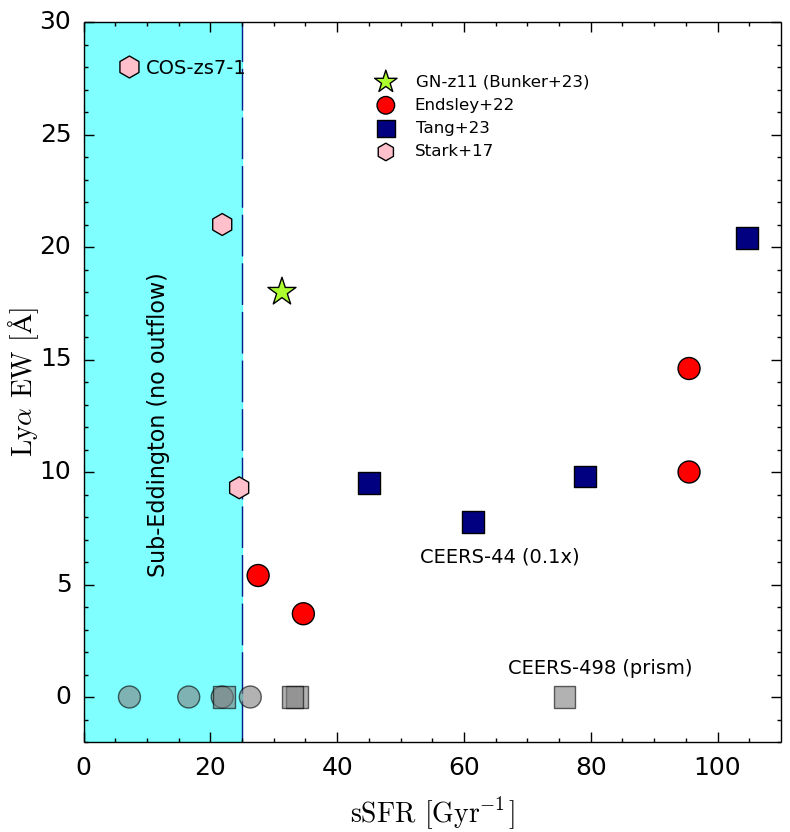}
\caption{Ly$\alpha$ EW vs. sSFR for 20 galaxies at $z\simgt 7$. The data points are taken from the following samples: \citet{Tang23} (blue squares), \citet{Endsley23} (red circles), \citet{Stark17} (pink hexagons), with grey points in each sample showing non-detections. Also shown is GN-z11 \citep[][green star]{Bunker23}. The EW of CEERS-44 has been multiplied by 0.1 for display purpose. The cyan band shows the sub-Eddington regime in which $\rm sSFR < 25\ Gyr^{-1}$ (no outflows expected). The two exceptions (CEERS-498 and COS-zs7-1) to our outflow condition are individually marked and discussed in the text.   
}
\label{Fig:EW_sSFR}
\end{figure}

\subsection{Ly$\alpha$ escape}\label{sec:Lya_escape}
The inferred sizes of the ionized bubbles carved by $z>10$ galaxies are actually too small\footnote{Alternatively, a larger ionized bubble could be created by the collective contribution of fainter objects clustered around the central, luminous one \citep{Castellano18, Tilvi20, Hu21}}  ($< 0.28$ pMpc, see Tab. \ref{Tab:properties}) to allow the escape of Ly$\alpha$ photons.  Outflows can significantly increase the visibility of the Ly$\alpha$ line from sources at high-$z$, where in principle the line should be strongly attenuated by the overwhelmingly neutral IGM \citep{Dijkstra10, Stark17, Mason18, Hayes23b, Bruton23}. 
Such facilitating role of outflows is supported by the evidence reported in Fig. \ref{Fig:EW_sSFR}, where along with GN-z11, we show the Lya EW vs. sSFR for three different samples \citep{Tang23, Endsley23, Stark17} of $z>7$ galaxies in which Ly$\alpha$ has been searched for (total of 20 galaxies). 

In the \citet{Tang23} sample we select 8 galaxies for which Ly$\alpha$ and stellar mass data are simultaneously available. Among these, 4 have been detected in Ly$\alpha$ (CEERS-1019, CEERS-1027, CEERS-698, CEERS-44). All of them  have velocity offsets $\simgt 300\, \kms$ and meet our super-Eddington condition, having $\rm sSFR = (79, 104, 45, 61)\, \rm Gyr^{-1} > sSFR^\star$. The remaining 4 non-detected galaxies all show low values of sSFR $\simlt 35\, \rm Gyr^{-1}$, with the only exception of CEERS-498 which however has only been observed vith the NIRSpec prism at low ($R\simeq 100$) resolution. Thus, Ly$\alpha$ could still be detected in a higher $R$ grating observation, similarly to the case of GN-z11 \citep{Bunker23}.  

Even more striking are the results by \citet{Endsley23}, who searched from Ly$\alpha$ emission from 8 galaxies at $z \approx 7$ selected from the ALMA REBELS survey \citep{Bouwens22a, Ferrara22a}. Using MMT/Binospec spectroscopic data, they detected broad, highly redshifted (mean offset with respect to [CII] line systemic redshift equal to $223\,\kms$) Ly$\alpha$ emission from 4 galaxies. These four systems have $\rm sSFR > sSFR^* = 25\, \rm Gyr^{-1}$.  On the contrary, the four non-detected REBELS sources fall in the sub-Eddington regime in which radiation-driven outflows are not expected. 

Finally, \citet{Stark17} report Ly$\alpha$ detection from three sources (EGS-zs8-1, EGS-zs8-2, COS-zs7-1). While two of them have a sSFR very close to the critical one, COS-zs7-1 has a low sSFR of $7.2\,\rm Gyr^{-1}$ and therefore it represents an exception to our criterion. Such anomaly can be explained by the fact that although outflows are clearly favoring Ly$\alpha$ escape, other properties, such as low \HI column densities, might also produce a large Ly$\alpha$ EW. 

In summary, though, we find that among the 12 Ly$\alpha$ detected sources, 11 have a sSFR compatible with the presence of an outflow; among the 8 non-detected sources, instead, 7 are in the sub-Eddington regime where the outflow should not develop. We note that the range $\rm 20< sSFR/Gyr^{-1} < 35$ is populated with both detections and non-detections (Fig. \ref{Fig:EW_sSFR}), likely as a result of the uncertainties of local variation of the $\rm sSFR^\star$ value, which in turn depends on the details of dust properties and radiative transfer (see Sec. \ref{sec:super-Edd_gals}). These results show that outflows play a paramount role for Ly$\alpha$ visibility.

\subsubsection{Ly$\alpha$ halos}
We conclude this Section with another potential prediction of our model. According to the discussion in Sec. \ref{Sec:outflows}, the terminal velocity $v_\infty$ is reached at the radius $r_{\tau=1} =2 (M_\star/10^9 \msun)^{1/2}\, \rm kpc$, where the outflow becomes optically thin and the velocity shift allows Ly$\alpha$ photons to escape. On this basis we expect super-Eddington galaxies to show an extended Ly$\alpha$ halo whose typical size is $\approx r_{\tau=1}$. Interestingly, \citet{Scholtz23} report the detection of a Ly$\alpha$ halo of size $0.8-3.2$ kpc around GN-z11. This value is comparable to the value  $r_{\tau=1} = 1.4$ kpc we predict for this system. Although a more refined modeling of the Ly$\alpha$ halo properties based on numerical simulations (see, e.g., \citealt{Hutter22}) is necessary to draw firm conclusions, it is nevertheless encouraging to see that the model successfully compares with the data.   

\subsection{LyC escape}\label{sec:LyC_escape}
Outflows might also facilitate the escape of LyC photons \citep{Hogarth20} by carving highly ionized channels in the galaxy. In fact, we have seen that GN-z11 Ly$\alpha$ modeling requires that the \HI fraction in the outflow region must be very small, $x_{\rm HI} = 4.2\times 10^{-3}$.  In addition, outflows also decrease the UV opacity by pushing dust on larger radial scales. The fact that the LyC escape fraction, $f_{\rm esc}$, increases with sSFR has been very convincingly demonstrated by \citet[][see their Fig. 17]{Flury22} using the Low-redshift Lyman Continuum Survey (LzLCS). They find that no $z \approx 0.3$ LyC emitters with $f_{\rm esc}> 10\%$ have $\rm sSFR < 10\ Gyr^{-1}$. 

Even more extreme is the case of the three galaxies (EGS-42501, EGS-41419, and EGS-47688) observed with {\it JWST}/NIRCam in the redshift range $7.2 < z < 8.4$ \citep{topping2022}. These exhibit both extremely blue UV slopes ($\beta \approx -3.1$), and very compact ($r_e <260$ pc) sizes\footnote{Similar conclusions are reached by \citet{Mascia23} from an analysis of 70 spectroscopically confirmed star-forming galaxies at  $6 < z < 9$ from the Cosmic Evolution Early Release Science (CEERS) survey.}. Noticeably, the LyC escape fraction for these three galaxies is very large ($f_{\rm esc}=0.8, 0.57, 0.65$); their sSFR is clearly super-Eddington, $\rm sSFR = (47, 99, 239)\ Gyr^{-1}$. 

From these evidences a coherent picture emerges in which the high sSFR produces radiatively-driven outflows that clear the dust (in turn determining steep UV spectral slopes), and carve ionized, dust-transparent channels though which LyC photons can easily leak.   

\subsection{Supernova explosions}
A key feature of high redshift galaxies is that the ISM of these systems is extremely dense, likely as a result of their very compact nature. In a recent study based on {\it JWST}/NIRSpec GLASS, ERO, and CEERS programs, \citet{Isobe23} found that the electron densities of  $z = 4-9$ galaxies are $n_e \simgt 300\ \cc$, i.e.  significantly higher than those of low-$z$ galaxies at a given stellar mass and sSFR. This finding is consistent with the determination of the gas electron density in GN-z11 ($z=10.6$) made by \citet{Senchyna23}. From the analysis of the NIV] line, they obtain $n_e > 10^5\ \cc$.  

Under such high density  conditions supernova explosions rapidly lose energy due to radiative losses (the so-called catastrophic cooling, \citealt[][]{Terlevich1992, Pizzati20}), and the shock is quickly (time scale of few $\times 100$ yr) damped into sound waves. Similar conclusions concerning the inefficiency of supernova-driven outflows in early massive galaxies are reached in \citet[]{Bassini22}.  Moreover,  \citet{Nath2022} find that in high-$z$  galaxies the gas ejected by supernova-driven shocks falls back onto the system.

A second important point is that in dense environments the radiation pressure, $p_r$, can greatly exceed the thermal pressure, $p_h$, sustained by SN explosions. Following \citet{McKee77, thompson2005}, and assuming a supernova explosion energy $E_0 = 10^{51}\ \rm erg$, a straightforward calculation shows that the ratio of the two pressures is
\be
\frac{p_h}{p_r} = 0.004 \left( \frac{100\ \rm pc}{r_e} \right) \left( \frac{n}{10^{3}\ \cc}\right)^{-4/7}.
\label{eq:p_ratio}
\ee
Thus, due their compactness and high density, radiation-pressure is the dominant dynamical factor ($\sim 250 \times$)  in super-early galaxies. Hence, the contribution from supernova explosions in driving outflows can be neglected to first order.

These arguments are further substantiated by the results in \citet{Gelli23b} who studied the star formation quenching mechanism in two high-$z$ galaxies recently detected by {\it JWST}, JADES-GS-z7-01-QU ($z=7.3$) and MACS0417-z5BBG ($z=5.2$). They find that SN feedback largely fails to reproduce the observed 
quenched SF history. They suggest instead that star formation is rapidly suppressed by radiation-driven dusty outflows sustained by the high 
sSFR ($43\, \rm Gyr^{-1}$ and $25\, \rm Gyr^{-1}$, respectively) of the two galaxies.

\subsection{The possible role of AGN}
{\it JWST} surveys are identifying a potentially large number of Active Galactic Nuclei (AGN) in super-early galaxies \citep{Fudamoto22a, Larson23, Goulding23, Maiolino23b, Fujimoto23b}. These identifications are mostly based on the presence of a broad component in some emission lines, typically but not exclusively in the Balmer series, with velocities of $400-1200\, \kms$. However, such detections are sometimes reported with low ($2-3\, \sigma$) significance.  Moreover, this approach might bias the observations toward sources whose broad component EW is high and/or whose width is sufficiently broad to be resolved with {\it JWST} instruments. 
In any case, as shown  in Sec. \ref{Sec:outflows}, comparable velocities can be achieved (eq. \ref{eq:terminal_v1}) also by radiation-driven outflows provided (i) the star formation efficiency, $\epsilon_\star$, is high, and/or (b) the ejected mass fraction, $f_M$, are low. 

When including also high ionization line detections and UV LF excess, \citet{Fujimoto23b} estimated that the AGN fraction at $z\ge 8.5$ is about $10-35$\%. On the other hand, \citet{Volonteri23} pointed out that the number density of {\it JWST} candidate galaxies far outnumbers that of the highest-$z$ quasar hosts, thus allowing for only $\simeq 1$ high-$z$ AGN every 1000 of these galaxies. 

Thus, together with the virtual lack of X-ray detections (but see \citealt{Bogdan23} for a tentative one), the actual presence of AGN in super-early galaxies remains an intriguing possibility. An additional problem is that the measured AGN UV LF would imply, for a standard AGN spectral template, an X-ray background $\approx 10\times$ higher than constrained by current experiments (Padnamabhan \& Loeb, in prep.). Hence, if present, super-early AGN should show a peculiarly low X-ray-to-UV emissivity. 

In terms of the stellar mass, an AGN should largely dominate the UV luminosity in order to affect such determination. \citet{Casey23} using the {\it JWST}/NIRCam COSMOS-Web survey to study the most luminous  ($-20.5 > M_{\rm UV} > -22$) $z\simgt 10$ candidates showed that this would imply a minimum black hole mass of $\approx 5\times 10^6 \ \msun$. Such masses would be unexpectedly large for the downward-revised stellar mass estimates of their host galaxies, $\simlt 10^9\ \msun$.

In the framework of our model, whether the observed UV luminosity driving the outflow has a contribution from an AGN, in addition to that of stars, does not make an actual difference. However, because in that case the same luminosity would correspond to a lower\footnote{As noted by \citet{D'Silva23}, including an AGN component when fitting spectral energy distributions the resulting cosmic star formation history is decreased by $\simeq 0.9$ dex.} SFR while the stellar mass remains unaltered, the $\rm sSFR^\star$ value derived in eq. \ref{eq:ssfr_thresh} should be interpreted as an upper limit to the sSFR necessary to drive the outflow.

\section{Summary}\label{sec:summary}
Radiation-driven outflows should become progressively more common towards high redshifts as a result of the increasing fraction of super-Eddington galaxies at early times. Outflows provide a coherent interpretation of {\it JWST} data showing (a) an unexpected excess of super-early, bright galaxies (b) with very blue colors, (c) some of which are able to transmit their Ly$\alpha$ line through the largely neutral IGM at $z>10$.  The main results of this study are:

\begin{itemize}
\item[{\color{red} $\blacksquare$}]  Independently of redshift, galaxies with a $\rm sSFR > sSFR^\star \simeq 25\ Gyr^{-1}$ become super-Eddington and launch an outflow that clears the dust making the galaxy brighter and bluer (Sec. \ref{sec:super-Edd_gals}). Such model successfully reproduces the observed  $z>10$ UV LF,  and the redshift evolution of the cosmic SFR density and sSFR (Sec. \ref{sec:sSFR}). 

\item[{\color{red} $\blacksquare$}] The fraction of galaxies that are in the super-Eddington regime increases with redshift (Fig. \ref{Fig:Super_Edd_frac}); it reaches  45\% (76\%) at $z=10 \, (14)$. 

\item[{\color{red} $\blacksquare$}] The outflow velocity is $\approx 830 (\epsilon_\star/f_M)^{1/2}\ \kms$, where $\epsilon_\star$ and $f_M$ are the star formation efficiency and the fraction of the \textit{halo gas} expelled in the outflow, respectively (eq. \ref{eq:terminal_v1}). The outflow model is consistent with the outflow measurements in 12 super-Eddington galaxies of the JADES sample, yielding an average value $f_M \approx 0.04$ (Fig. \ref{Fig:carniani}), roughly corresponding to the entire galaxy gas mass.

\item[{\color{red} $\blacksquare$}] Outflows can dramatically enhance the Ly$\alpha$ line transmission from high-$z$ sources even if the intergalactic medium (IGM) is almost neutral, as expected at $z\simgt 10$. The observed Ly$\alpha$ properties in GN-z11 are simultaneously recovered (Fig. \ref{Fig:Lya_profiles}) by the outflow model if $\log N_{\rm HI} \simeq 20.1$, implying that the outflow is largely ionized. We make analogous predictions for the Ly$\alpha$ visibility of other super-early galaxies (Tab. \ref{Tab:properties}). 

\item[{\color{red} $\blacksquare$}] We also compare the model with available Ly$\alpha$ surveys at $z>7$, finding that essentially all super-Eddington (sub-Eddington) galaxies are (not) detected in Ly$\alpha$ (Sec. \ref{sec:Lya_escape}). Super-Eddington galaxies might also feature Ly$\alpha$ halos, as observed in GN-z11. 

\item[{\color{red} $\blacksquare$}] Based on available low- and high-$z$ data, we show that the sSFR positively correlates with LyC escape fraction (Sec. \ref{sec:LyC_escape}). This is because a high sSFR produces radiation-driven outflows that clear the dust (in turn determining steep UV spectral slopes), and carve ionized, dust-transparent channels through which LyC photons leak. 

We conclude that radiation-driven outflows are a key factor to understand the properties and evolution of super-early galaxies.
\end{itemize}

\section*{Data Availability}
Data available on request.

\acknowledgments
We thank P. Arrabal-Haro, R. Bouwens, F. Cullen and D. McLeod and
M. Stefanon  for providing data and useful information in advance of
publication; M. Brada{\v{c}}, A. Calabr\'o, S. Carniani, M. Castellano, A. Dekel, V. D'Odorico, F. Fiore, Y. Harikane, A. Inoue, A. Loeb, D. McLeod, M. Ouchi, A. Pallottini, L. Pentericci, H. Padmanabhan, M. Stiavelli, S. Tacchella for useful discussions and comments.
Support from the ERC Advanced Grant INTERSTELLAR H2020/740120 is kindly acknowledged. This research is also generously supported by a Carl Friedrich von Siemens-Forschungspreis der Alexander von Humboldt-Stiftung Research Award. 
Plots in this paper produced with the \textsc{matplotlib} \citep{Hunter07} package for \textsc{PYTHON}.    

\bibliographystyle{aasjournal}
\bibliography{paper}



\end{document}